\documentclass[iop,apj,tighten]{emulateapj}
\usepackage{mathptmx}

\shorttitle{SOLAR DOPPLER SHIFT, INTENSITY, AND DENSITY
  OSCILLATIONS}

\shortauthors{MARISKA and MUGLACH}

\slugcomment{Accepted 2010 February 19}

\begin{document}

\title{DOPPLER SHIFT, INTENSITY, AND DENSITY OSCILLATIONS
  OBSERVED WITH THE EUV IMAGING SPECTROMETER ON \textit{HINODE}}

\author{John T. Mariska and K. Muglach\altaffilmark{1}}
\affil{Space Science Division, Code 7673, Naval Research
  Laboratory, Washington, DC 20375; mariska@nrl.navy.mil}
\altaffiltext{1}{Also at Artep Inc., Ellicott City, MD 21042}

\begin{abstract}
Low-amplitude Doppler-shift oscillations have been observed in
coronal emission lines in a number of active regions with the EUV
Imaging Spectrometer (EIS) on the \textit{Hinode} satellite. Both
standing and propagating waves have been detected and many
periods have been observed, but a clear picture of all the wave
modes that might be associated with active regions has not yet
emerged. In this study, we examine additional observations
obtained with EIS in plage near an active region on 2007 August
22--23. We find Doppler-shift oscillations with amplitudes
between 1 and 2 km~s$^{-1}$ in emission lines ranging from
\ion{Fe}{11} 188.23~\AA, which is formed at $\log T = 6.07$ to
\ion{Fe}{15} 284.16~\AA, which is formed at $\log T = 6.32$.
Typical periods are near 10 minutes. We also observe intensity
and density oscillations for some of the detected Doppler-shift
oscillations. In the better-observed cases, the oscillations are
consistent with upwardly propagating slow magnetoacoustic waves.
Simultaneous observations of the \ion{Ca}{2} H line with the
\textit{Hinode} Solar Optical Telescope Broadband Filter Imager
show some evidence for 10-minute oscillations as well.
\end{abstract}


\section{INTRODUCTION}
\label{intro}

Because the outer regions of the solar atmosphere are threaded by
a magnetic field, they can support a wide range of oscillatory
phenomena. Theoretical aspects of these waves and oscillations
have been the subject of extensive investigations
\citep[e.g.,][]{Roberts1983,Roberts1984}. These initial studies
were mainly driven by radio observations of short period
oscillations \citep[e.g.,][]{Rosenberg1970}. More recent
detections of coronal oscillatory phenomena have resulted in
considerable additional theoretical work. Recent reviews include
\citet{Roberts2000} and \citet{Roberts2003}. Observational
detections of coronal oscillatory phenomena include detections of
spatial oscillations of coronal structures, which have been
interpreted as fast kink mode MHD disturbances
\citep[e.g.,][]{Aschwanden1999}; intensity oscillations, which
have been interpreted as propagating slow magnetoacoustic waves
\citep[e.g.,][]{DeForest1998,Berghmans1999}; and Doppler shift
oscillations, which have been interpreted as slow mode MHD waves
\citep[e.g.,][]{Wang2002}. The interaction between the
theoretical work and the growing body of oscillation observations
provides fertile ground for testing our understanding of the
structure and dynamics of the corona.

The EUV Imaging Spectrometer (EIS) on the \textit{Hinode}
satellite is an excellent tool for studying oscillatory phenomena
in the corona. \citet{Culhane2007} provides a detailed
description of EIS, and the overall \textit{Hinode} mission is
described in \citet{Kosugi2007}. Briefly, EIS produces stigmatic
spectra in two 40~\AA\ wavelength bands centered at 195 and
270~\AA. Two slits (1\arcsec\ and 2\arcsec) provide line
profiles, and two slots (40\arcsec\ and 266\arcsec) provide
monochromatic images. Moving a fine mirror mechanism allows EIS
to build up spectroheliograms in selected emission lines by
rastering a region of interest. With typical exposure times of 30
to 90~s, however, it can take considerable time to construct an
image. Higher time cadences can be achieved by keeping the EIS
slit or slot fixed on the Sun and making repeated exposures. This
sit-and-stare mode is ideal for searching for oscillatory
phenomena.

EIS Doppler shift data have already been used for a number of
investigations of oscillatory phenomena.
\citet{VanDoorsselaere2008} have detected kink mode MHD
oscillations with a period near 5 minutes. \citet{Mariska2008}
observed damped slow magnetoacoustic standing waves with periods
of about 35 minutes. \citet{Wang2009} have detected slow mode
magnetoacoustic waves with 5 minute periods propagating upward
from the chromosphere to the corona in an active region.
\citet{Wang2009a} have also observed propagating slow
magnetoacoustic waves with periods of 12 to 25 minutes in a large
fan structure associated with an active region. Analysis of
oscillatory data with EIS is still just beginning. The amplitudes
observed have all been very small---typically 1 to 2 km~s$^{-1}$.
Thus a clear picture of the nature of the low-amplitude coronal
oscillations has yet to emerge. In this paper, we add to that
picture by analyzing portions of an EIS sit-and-stare active
region observation that shows evidence for Doppler-shift
oscillations in a number of EUV emission lines. We also use data
from the \textit{Hinode} Solar Optical Telescope (SOT) to relate
the phenomena observed in the corona with EIS to chromospheric
features and the magnetic field.

\section{OBSERVATIONS}

\begin{figure*}
\plotone{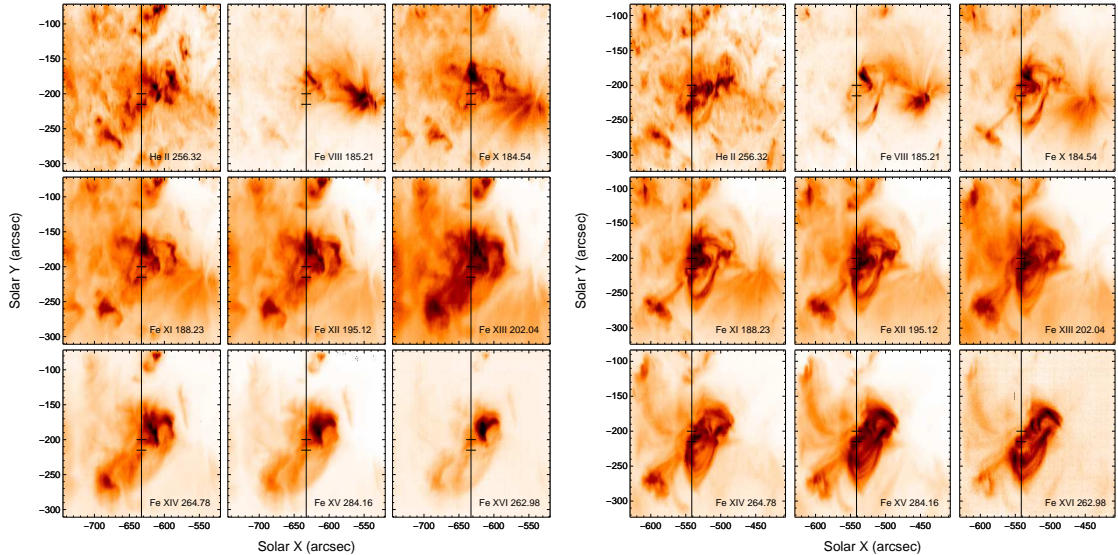} 
\caption{Context EIS spectroheliograms in 9 wavelength windows
  obtained from 13:33:46 to 17:56:57 UT on 2007 August 22 (left)
  and from 01:55:43 to 06:18:53 UT on 2007 August 23 (right). The
  location of the EIS slit for the sit-and-stare observation is
  marked with the vertical line. The horizontal lines along the
  slit mark the location of the oscillation data analyzed in this
  paper. The post-flare loops discussed in the text are the
  bright features east and south of the marked region in the
  three bottom right panels.}
\label{fig:context_panel}
\end{figure*}

The observations discussed in this paper were taken in an area of
enhanced EUV and soft X-ray emission just west of NOAA active
region 10969. The complete EIS data set consists of a set of
$256\arcsec \times 256\arcsec$ context spectroheliograms in 20
spectral windows, followed by 7.4~h of sit-and-stare observations
in 20 spectral windows, and finally a second set of context
spectroheliograms. All the EIS sit-and-stare data were processed
using software provided by the EIS team. This processing removes
detector bias and dark current, hot pixels, dusty pixels, and
cosmic rays, and then applies a calibration based on the
prelaunch absolute calibration. The result is a set of
intensities in ergs cm$^{-2}$ s$^{-1}$ sr$^{-1}$ \AA$^{-1}$. The
data were also corrected for the EIS slit tilt and the orbital
variation in the line centroids. For emission lines in selected
wavelength windows, the sit-and-stare data were fitted with
Gaussian line profiles plus a background, providing the total
intensity, location of the line center, and the width.

Figure~\ref{fig:context_panel} shows the pre- and
post-sit-and-stare spectroheliograms and provides a more detailed
view of the area on the Sun covered by the sit-and-stare
observations. The spectroheliograms were obtained with the
1\arcsec\ EIS slit using an exposure time of 60~s. Note that the
two sets of spectroheliograms have small differences in the
pointing. Examination of the two sets of images shows that
considerable evolution has taken place in the observed region
between the times each set was taken. In particular, new loop
structures have developed at locations to the south of the bright
core of the emitting region, suggesting that some flare-like
heating may have taken place between the two sets of context
observations. Space Weather Prediction Center data show that
there was one B1.2 class flare between 07:50 and 08:10~UT on 2007
August 22 at an unknown location. Since other flares occurred in
this active region, it is likely that this one is also associated
with it. There were, however, no events recorded during the time
of the EIS observations.

\begin{figure}[b]
\plotone{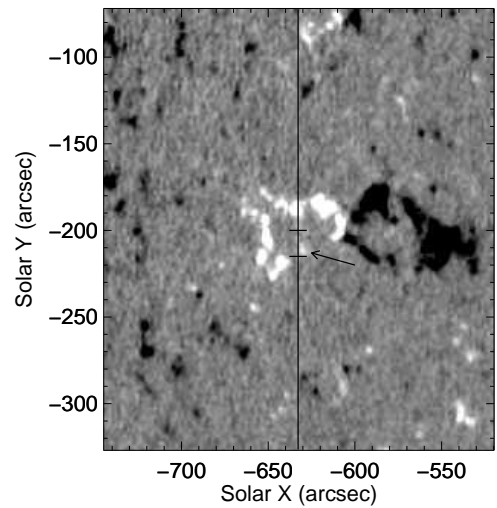} 
\caption{A portion of an MDI magnetogram taken on 2007 August 22
  at 14:27 UT showing the area covered by the EIS
  spectroheliograms and the approximate location of the EIS slit
  during the sit-and-stare observation. The horizontal lines
  along the slit mark the location of the oscillation data
  analyzed in this paper, and the area of weak magnetic flux that
  is the focus of this analysis is marked with an arrow. The image
  has been scaled so that the range of magnetic fluxes displayed
  is $\pm100$ Gauss.}
\label{fig:mdi_mag}
\end{figure}

\begin{figure}
\plotone{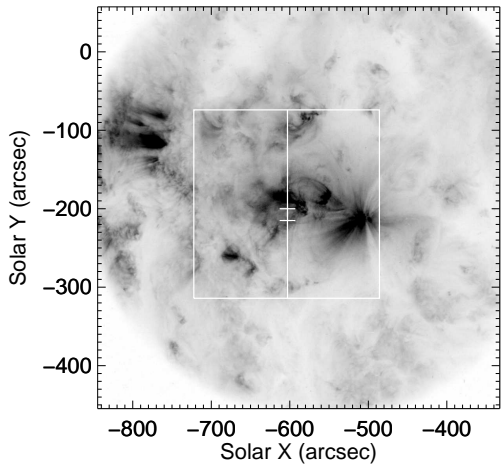} 
\caption{\textit{TRACE} 171~\AA\ (Fe$\;$\small{IX/X}) image taken
  on 2007 August 22 at 20:01:19 UT. The box shows the location of
  the EIS spectroheliograms taken before and after the
  sit-and-stare observations, and the vertical line with the box
  shows the location of the sit-and stare observation. The
  horizontal lines along the slit mark the location of the
  oscillation data analyzed in this paper.}
\label{fig:trace}
\end{figure}

Both magnetogram data from the \textit{SOHO} Michelson Doppler
Imager (MDI) and \textit{Hinode} Solar Optical Telescope (SOT)
and coronal images taken in the 171 \AA\ (Fe$\;$\small{IX/X})
filter with the \textit{Transition Region and Coronal Explorer}
(\textit{TRACE}) were also examined. The magnetograms, one of
which is shown in Figure~\ref{fig:mdi_mag} with the location of
the EIS slit for the sit-and-stare observation indicated, show an
extended bipolar plage region without any visible sunspots. As
can be seen in the magnetogram, the EIS slit crosses two regions
of positive (white) polarity. The weaker one, which has a size of
about 6\arcsec and is marked with an arrow, is the focus of this
analysis.

\begin{figure*}
\plotone{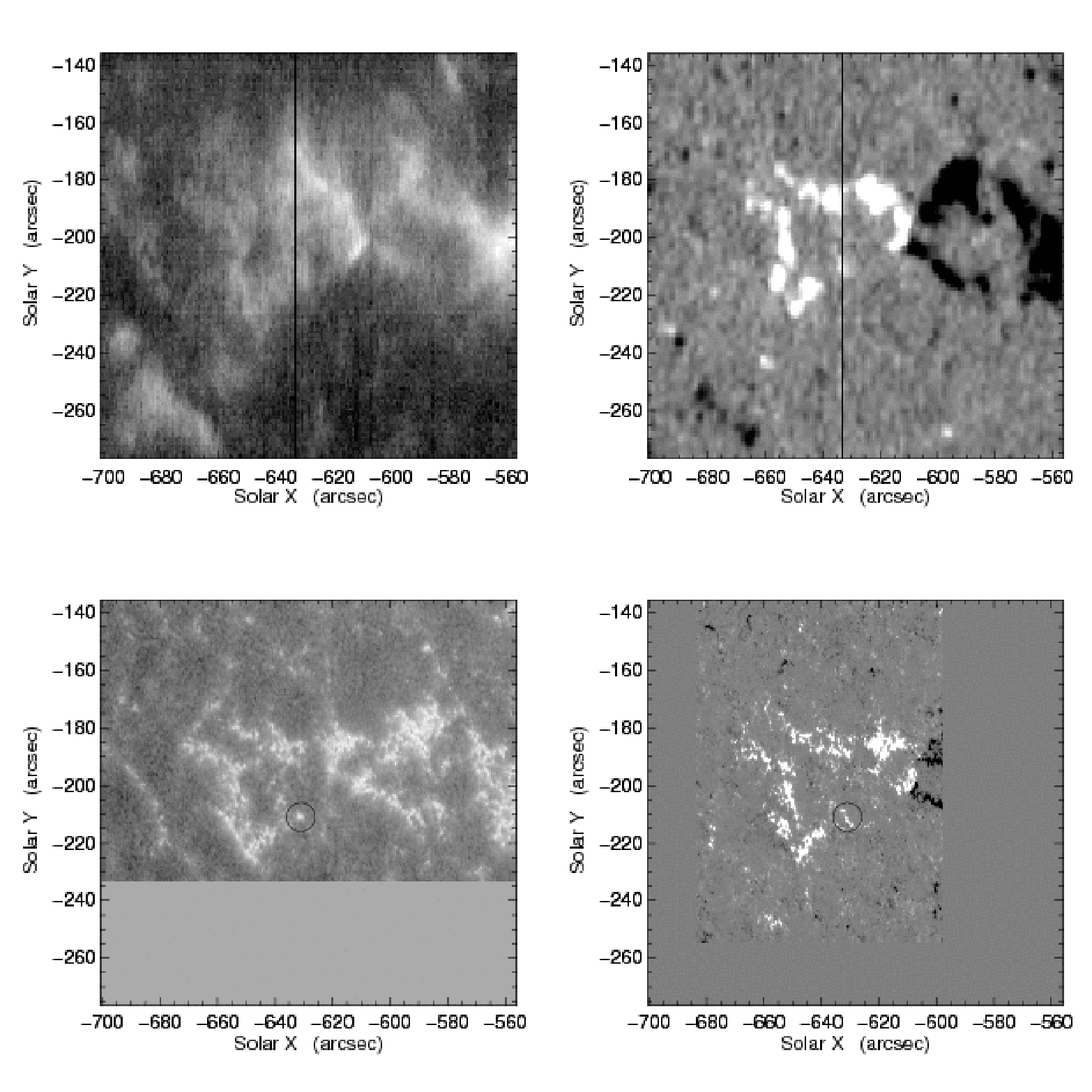} 
\caption{Coalignment of EIS and SOT. Top left: EIS
  spectroheliogram taken in \ion{O}{6} 184.12 \AA\ at the same
  time as the spectroheliograms shown in
  Figure~\ref{fig:context_panel}, the dark vertical line near the
  center marks the slit location for the sit-and-stare
  observations. Top right: MDI full disk magnetogram taken at
  14:27 UT. Bottom right: \textit{Hinode} SP magnetogram, taken
  between 18:03 UT and 19:01 UT, at the same time the \ion{Ca}{2}
  H time sequence was taken. Bottom left: temporal average of the
  time sequence in \ion{Ca}{2} H taken between 19:50 and 21:40
  UT. The black circles in the bottom images encompass a small
  network bright point that overlaps with the EIS slit.}
\label{fig:coalign}
\end{figure*}

The \textit{TRACE} images, one of which is shown in
Figure~\ref{fig:trace}, show that the western part of the region
near the right edge of the box that outlines the edges of the
second EIS spectroheliogram exhibits fan-like coronal structures
extending radially out in all directions, while the eastern part
consists of shorter loop structures that connect the opposite
polarities. The full set of \textit{TRACE} images that cover most
of the time period covered by the context and sit-and-stare
observations are included as a movie. The movie shows generally
quiescent behavior in the fan-like structures, but multiple
brightenings in the eastern loop system. The largest of these
took place at around 01:31~UT on August 23 and shows a filament
eruption, which resulted in a system of post-flare loops that are
visible in the \ion{Fe}{15} 284.16~\AA\ and \ion{Fe}{16}
262.98~\AA\ spectroheliograms in the right panels of
Figure~\ref{fig:context_panel}. Some of the loops in the eastern
part cross the EIS slit location for the sit-and-stare
observations, but they are for the most part north of the area we
focus on.

We have also examined images taken with the Extreme Ultraviolet
Imaging Telescope \citep[EUVI,][]{Wuelser2004} on the Ahead (A)
and Behind (B) spacecraft of the \textit{Solar Terrestrial
  Relations Observatory} (\textit{STEREO}). EUVI observes the
entire solar disk and the corona up to 1.4 R$_\sun$ with a pixel
size of 1.59\arcsec. On 2007 August 22, the separation angle of
\textit{STEREO} with Earth was 14.985$\degr$ for spacecraft A and
11.442$\degr$ for spacecraft B. Our target region was rather near
the limb as seen from spacecraft A. EUVI/B, however, provided
continuous on-disk images between 2007 August 22, 11:00 UT and
2007 August 23, 11:30 UT, which allowed us to study the
development of the coronal structures of the region during the
time of the EIS observations.

We produced movies in all four EUVI channels (171~\AA , 195~\AA,
284~\AA, and 304~\AA), with 171~\AA\ having the highest cadence
of 2.5 min. The EUVI movies generally show the same behavior as
that shown in the \textit{TRACE} data. EUVI captured two larger
events during this period, the first one started at around
12:24~UT on August 22, showing multiple brightenings of the
eastern loop system although it is not clear if a CME had been
launched. The second one took place at around 01:31~UT on August
23, involving the western part of the region and shows a filament
eruption with a system of post-flare loops evolving at the time
the second EIS context spectroheliogram was taken.

SOT obtained \ion{Ca}{2} H observations at a one-minute cadence
during the time interval of the sit-and-stare observations that
are the focus of this study. To coalign those observations we use
an EIS \ion{O}{6} 184.12 \AA\ spectroheliogram, MDI magnetograms,
SOT Spectro-Polarimeter (SP) magnetograms, and SOT \ion{Ca}{2} H
line filtergrams. The EIS \ion{O}{6} 184.12 \AA\ spectroheliogram
was part of the scan shown in Figure~\ref{fig:context_panel}
(left), taken several hours before the sit-and-stare
observations. Coalignment was carried out by rescaling the
images, using cross-correlation techniques to overlap the
selected subfields and finally blinking the images to ensure
minimal offsets.

Figure~\ref{fig:coalign} displays the \ion{O}{6} 184.12
\AA\ spectroheliogram in the top left corner, with the dark
vertical line representing the slit position of the sit-and-stare
observations. The top right image is from a full-disk MDI
magnetogram taken at 14:27 UT. The three small structures in the
lower left corner of the magnetogram were used to coalign the
\ion{O}{6} image, which shows bright emission structures at the
same locations. In the next step we use SP data to coalign with
the MDI magnetogram. The SP was scanning the target region on
2007 August 22 between 18:03 UT and 19:01 UT, at the same time
the \ion{Ca}{2} H image sequence was taken. The lower right image
in Figure~\ref{fig:coalign} gives the apparent longitudinal field
strength as derived from the Stokes spectra. Comparing the two
magnetograms we can see that the overall structure of the
bi-polar region has hardly evolved over time and that the SP data
show considerable additional fine structure in the magnetic
field. We also note that in the time between the MDI magnetogram
and SP magnetogram flux cancellation has taken place at the
polarity inversion line of the bi-polar region (at around $x =
-610$\arcsec). The eruption observed a few hours later in EUVI
and \textit{TRACE} was probably triggered by this flux
cancellation.

\begin{figure}
\plotone{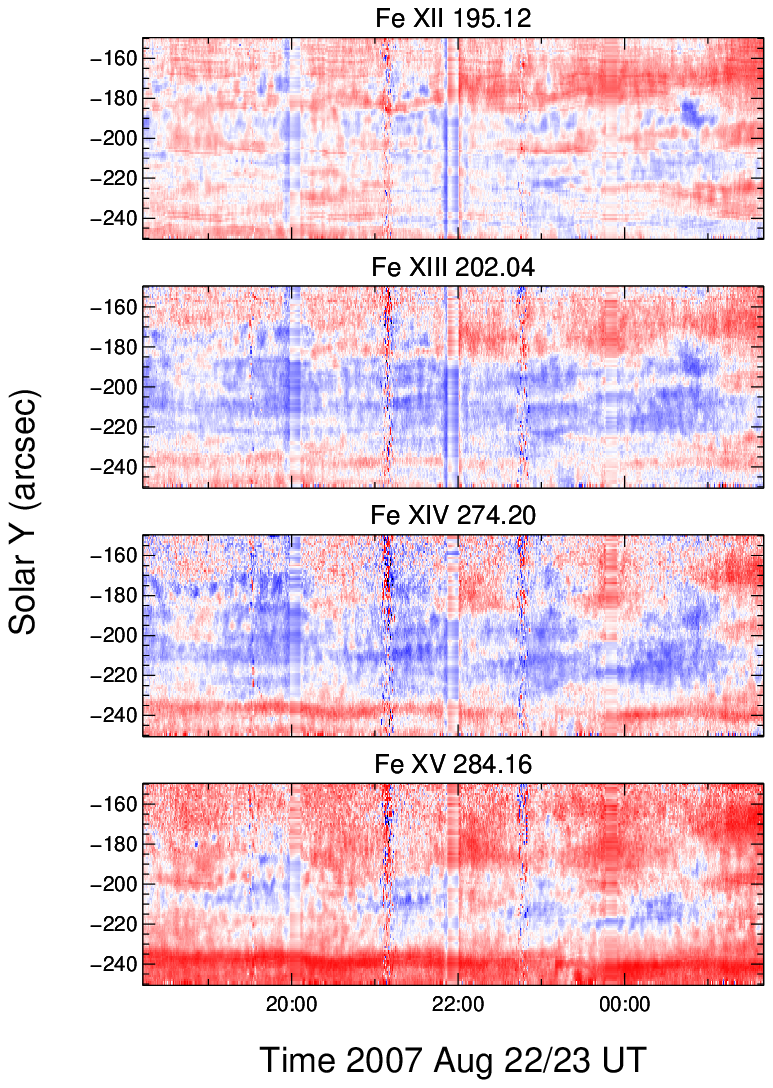} 
\caption{EIS sit-and-stare Doppler-shift data in four emission
  lines covering the entire sit-and-stare observation. The
  Doppler shift in each window has been adjusted so that the zero
  value is the average over the window. The maximum and minimum
  values plotted in each case are $+20$ and $-20$ km s$^{-1}$,
  respectively. This study focuses on the spatial range from
  $-200$\arcsec\ to $-215$\arcsec\ in the time interval from
  20:07 to 21:53~UT.}
\label{fig:doppler_local}
\end{figure}

\begin{figure}
\plotone{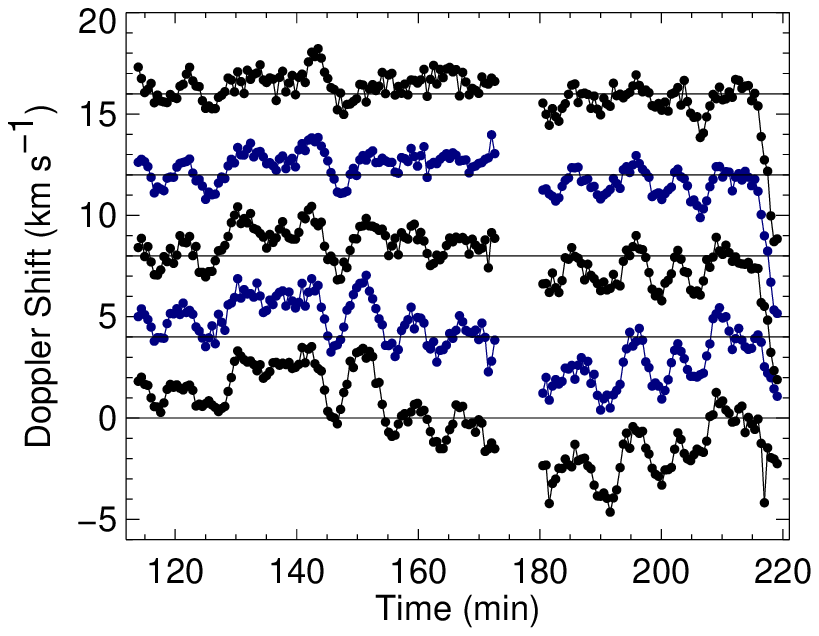} 
\caption{Doppler shift data averaged over the 16 detector rows
  from $-200$\arcsec\ to $-215$\arcsec. The emission lines shown
  from top to bottom are \ion{Fe}{11} 188.23~\AA, \ion{Fe}{12}
  195.12~\AA, \ion{Fe}{13} 202.04~\AA, \ion{Fe}{14} 274.20~\AA,
  and \ion{Fe}{15} 284.16~\AA. Time is measured in minutes from
  the start of the sit-and-stare data at 18:13:06 UT. The zero
  value for Doppler shifts is set to the average Doppler shift in
  each line and is shown using the horizontal lines. Data for
  each emission line are displaced from the next set by 4 km
  s$^{-1}$.}
\label{fig:doppler_average}
\end{figure}

The target area for the oscillation study is in the black circle,
a small magnetic region of positive polarity. In the final
coalignment step the SOT \ion{Ca}{2} H images were coaligned with
the SP scan. The bottom left image in Figure~\ref{fig:coalign}
shows the temporal average of the time sequence taken between
19:50 UT and 21:40 UT. The black circle encompasses a small
network bright point that overlaps with the EIS slit and the
region averaged along the slit in the analysis that is the focus
of this paper. As we have already noted, the EIS
spectroheliograms taken before and after the sit-and-stare
observation show that considerable structural evolution has taken
place. In the \ion{Fe}{15} 264.78~\AA\ spectroheliogram taken
after the sit-and-stare observation, there is significant
emission within the 6\arcsec\ region of the EIS slit that is the
focus of this paper, suggesting that the location is near one end
of one of the coronal loops seen in the lower three panels in the
right side of Figure~\ref{fig:context_panel}. On the other hand
the images in the EIS spectroheliograms taken immediately before
the sit-and-stare observation show much less evidence for a loop
rooted at that location. Thus we can only conclude that the
bright feature we observe in the \ion{Ca}{2} H line may be the
footpoint of a coronal loop. We do note, however, that there are
no other strong magnetic concentrations close to the location of
the region that is the focus of this study. The ones to the
south-east are about 10\arcsec\ away---more than our co-alignment
errors. We believe that loop footpoints will be anchored in
strong flux concentrations.

The basic building block for the sit-and-stare observation was
EIS Study ID~56. This study takes 50 30 s exposures in 20
spectral windows with a solar $y$-position coverage of
400\arcsec. The entire sit-and-stare observation consisted of
four executions of the study, with each execution invoked with a
repetition count of four. Thus, the full sit-and-stare data set
consists of 16 repetitions of the basic building block. There was
a brief delay between each invocation of the study, leading to
gaps in the resulting time series.

Over both orbital periods and shorter time intervals, the
\textit{Hinode} pointing fluctuates in both the solar $x$- and
$y$-directions by up to about 3\arcsec\ peak-to-peak. This is due
to both spacecraft pointing variations and changes in the
location of EIS relative to the location of the Sun sensors on
the spacecraft. Studies of the latter variations have shown that
they are well-correlated with similar variations observed in data
from the \textit{Hinode} X-Ray Telescope (XRT). We have therefore
used a modified version of the software provided by the XRT team
to compute the average $y$-position of each pixel along the slit
over each exposure and then interpolated the fitted centroid
positions onto a uniform $y$-position grid.

It is not, of course, possible to correct the sit-and-stare
observations for fluctuations in the $x$-position on the Sun.
Plots of the $x$-position pointing fluctuations show that they
tend to be smoother than the $y$-direction fluctuations, and that
they are dominated by the orbital period. If periodic Doppler
shifts are present only in very small structures, we would thus
expect the signal to show a modulation with the orbital period.
Larger structures, $\geq 3$\arcsec\ in the $x$-direction, that
display coherent Doppler shifts should not be affected by the
spacecraft $x$-position pointing variations.

Over much of the EIS slit during the sit-and-stare observation
there is no evidence for interesting dynamical behavior in the
Doppler shift observations. The portion of the slit that covers
the brighter core area of the region does, however, show evidence
for periodic changes in the Doppler shifts.
Figure~\ref{fig:doppler_local} shows the measured Doppler-shift
data in four emission lines over this $y$-position range as a
function of time. The gaps between each invocation of the study
appear as wider pixels near 20, 22, and 0 UT. Note that the data
are also affected by passage of the spacecraft through the South
Atlantic Anomaly (SAA). A small region of compromised data
appears near 19:30 UT, and major SAA passages are evident near
21:00 and 22:45 UT.

The display shows clear evidence for periodic fluctuations in all
the emission lines at $y$-positions of roughly $-170$\arcsec\ to
$-180$\arcsec during the first hour shown in the plots. Periodic
fluctuations are also visible over $y$-locations centered near
$-210$\arcsec. Note that there is some evidence of longer period
fluctuations, for example in the \ion{Fe}{15}
284.16~\AA\ emission line. These fluctuations are probably the
result of the correction for orbital line centroid shifts not
fully removing those variations. In the remainder of this paper,
we focus on the area near $-210$\arcsec\ that shows evidence for
oscillatory phenomena.

\section{ANALYSIS}

\subsection{Doppler Shift Oscillations}

As we noted earlier, solar $y$-positions between
$-200$\arcsec\ and $-215$\arcsec\ show considerable oscillatory
behavior, particularly in the set of data taken beginning at
20:07:01 UT. Figure~\ref{fig:doppler_average} shows the averaged
Doppler shift data over the time period of this set of
observations for, from top to bottom, \ion{Fe}{11} 188.23~\AA,
\ion{Fe}{12} 195.12~\AA, \ion{Fe}{13} 202.04~\AA, \ion{Fe}{14}
274.20~\AA, and \ion{Fe}{15} 284.16~\AA. The Doppler shifts have
been averaged over the 16 detector rows from $-200$\arcsec\ to
$-215$\arcsec. Data from 172 to 180 minutes were taken during SAA
passage and have been removed from the plot.

The Doppler shift data over this portion of the EIS slit show
clear evidence for low-amplitude, roughly 2--4 km s$^{-1}$,
oscillatory behavior with a period near 10 minutes. For some of
the time period, particularly after 180 minutes, there appears to
be a clear trend for the oscillations to display increasing
amplitude as a function of increasing temperature of line
formation.

\begin{deluxetable}{lcccccc}
\tablecaption{Periods and Amplitudes Detected in Doppler Shift
  and Intensity Data\label{table:doppler_periods}}
\tablewidth{0pt} \tablecolumns{4} \tablehead{ \colhead{} &
  \colhead{Wavelength} & \colhead{Log $T$} & \colhead{$P_D$} &
  \colhead{$\delta v$} & \colhead{$P_I$} & \colhead{$\delta I/I$}
  \\ \colhead{Ion} & \colhead{(\AA)} & \colhead{(K)} &
  \colhead{(minutes)} & \colhead{(km s$^{-1}$)} &
  \colhead{(minutes)} & \colhead{(\%)} } \startdata \ion{Fe}{11}
& 188.23 & 6.07 & 10.0 & 1.2 & 11.8 & 1.1 \\ \ion{Fe}{12} &
195.12 & 6.11 & 10.1 & 1.1 & 11.4 & 0.9 \\ \ion{Fe}{13} & 202.04
& 6.20 & \phn 9.1 & 1.3 & 11.2 & 1.4 \\ \ion{Fe}{14} & 274.20 &
6.28 & \phn 9.1 & 1.3 & 11.4 & 2.0 \\ \ion{Fe}{15} & 284.16 &
6.32 & \phn 9.0 & 1.4 & \phn 7.6 & 2.4 \\ \enddata
\end{deluxetable}

\begin{figure}
\plotone{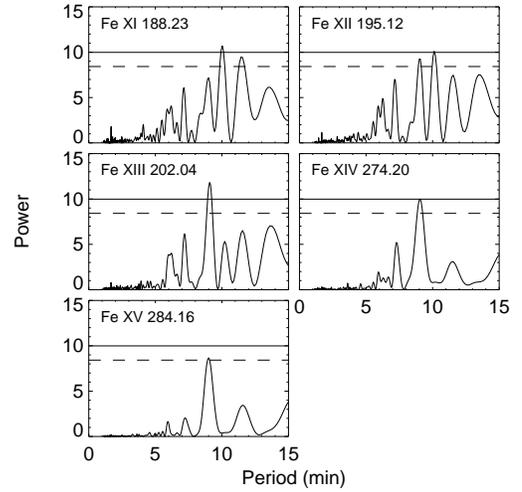} 
\caption{Periodograms computed for the Doppler shift data shown
  in Figure~\ref{fig:doppler_average}. The solid horizontal line
  on each plot indicates the power equivalent to a calculated
  false alarm probability of 1\% and the dashed line is for a
  false alarm probability of 5\%.}
\label{fig:pgram_v}
\end{figure}

Because there is a significant gap in the Doppler shift data,
neither Fourier time series analysis nor wavelet analysis is
appropriate. Instead, we examine the time series by calculating
periodograms using the approach outlined in \citet{Horne1986} and
\citet{Press1989}. Figure~\ref{fig:pgram_v} shows the
periodograms calculated from the Doppler shift data shown in
Figure~\ref{fig:doppler_average}. Also plotted on the figure are
the 99\% and 95\% significance levels. All but one of the time
series show a peak in the periodogram at the 99\% confidence
level, and the largest peak in the \ion{Fe}{15}
284.16~\AA\ emission line data is at the 95\% confidence level.
Table~\ref{table:doppler_periods} lists the period of the most
significant peak in each of the panels shown in the figure.

To estimate the amplitude of the oscillations, we detrend the
Doppler shift time series by subtracting a background computed
using averaged data in a 10-minute window centered on each data
point and then compute the rms amplitude as the standard
deviation of the mean. For a sine wave, the peak velocity is the
rms value multiplied by $\sqrt 2$. These peak velocity values are
also listed in the table in the $\delta v$ column. Visual
inspection of the data suggests that the numbers in the table are
smaller than what might be obtained by fitting the data. The
numbers do confirm the impression that the oscillation amplitude
increases with increasing temperature of line formation.

It is clear from the Doppler shift plots in
Figure~\ref{fig:doppler_average} that the oscillations are not
always present. Instead, they appear for a few periods and then
disappear. To understand better this behavior, we have fitted the
time intervals where oscillations are obvious with a combination
of a damped sine wave and a polynomial background. Thus, for each
time period where oscillations are present we assume that the
data can be fitted with a function of the form
\begin{equation}
v(t) = A_0 \sin(\omega t + \phi) \exp(-\lambda t) + B(t),
\end{equation}
where
\begin{equation}
B(t) = b_0 + b_1 t + b_2 t^2 + b_3 t^3 + \cdots
\end{equation}
is the trend in the background data. Time is measured from an
initial time $t_0$, which is different for each set of
oscillations we fit. The fits were carried out using
Levenberg-Marquardt least-squares minimization
\citep{Bevington1969}. Generally only two terms in the background
polynomial were necessary.

\begin{figure}
\plotone{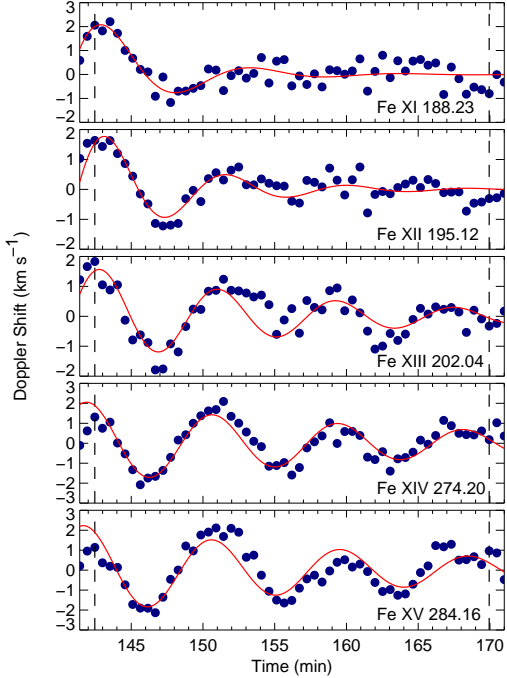} 
\caption{Decaying sine wave fits to the Doppler shift data
  beginning roughly 142 minutes after the start of the
  sit-and-stare observation. The plotted data has had the
  polynomial background removed. Vertical dashed lines show the
  time range used in the fitting.}
\label{fig:fit_fig_v}
\end{figure}

Figure~\ref{fig:fit_fig_v} shows the results of this fitting for
the Doppler shift data beginning 142.5 minutes after the start
of the sit-and-stare observation. All the fits show roughly the
same amplitudes, periods, and phases. Emission lines formed at
the higher temperatures (\ion{Fe}{14} and \ion{Fe}{15}) show
clear evidence for more than one full oscillation period. At lower
temperatures, the oscillatory signal damps much more rapidly.

Table~\ref{table:oscillation_d_fits} lists the amplitudes, $A_0$,
periods, $P$, phases, $\phi$, and the inverse of the decay rate,
$\lambda$, that result from fitting all the time periods in the
data for which a reasonable fit to the Doppler shift data could
be obtained. The periods are consistent with the results of the
periodogram analysis for the entire time interval. Generally, the
amplitudes are larger than the $\delta v$ values show in
Table~\ref{table:doppler_periods}. Note that some of the fits
show negative decay times, indicating that some of the
oscillations show a tendency to grow with time. In these cases,
this is not followed by a decay, but rather a rapid loss of the
oscillatory signal.

All the fits use the start times $t_0$ listed in the table and
thus the phase values for each time interval can not be directly
compared. When the phases are adjusted to a common start time,
the values do not agree. Thus while the periods are similar for
each time interval in which oscillations are observed, it appears
that each event is being independently excited.

\begin{deluxetable}{llcccc}
\tablecaption{Doppler Shift Oscillation
  Properties\label{table:oscillation_d_fits}} \tablewidth{0pt}
\tablehead{
  \colhead{$t_0$} & \colhead{} &
  \colhead{$A_0$} & \colhead{$P$} & \colhead{$\phi$} &
  \colhead{$\lambda^{-1}$} \\
  \colhead{(min)} & \colhead{Ion} & \colhead{(km s$^{-1}$)} &
  \colhead{(min)} & \colhead{(rad)} & \colhead{(min)} 
} \startdata
113.9 & \ion{Fe}{11} & 0.9 & \phn 8.8 & 1.8 & \phs \phn \phn 19.0 \\
113.9 & \ion{Fe}{12} & 0.7 & \phn 9.2 & 2.3 & $-216$ \\
113.9 & \ion{Fe}{13} & 0.7 & \phn 8.8 & 2.3 & \phn \phn $-73.3$ \\
113.9 & \ion{Fe}{14} & 0.3 & 10.7 & 3.9 & \phn \phn $-10.9$ \\
113.9 & \ion{Fe}{15} & 0.5 & \phn 9.5 & 2.9 & \phn \phn $-24.5$ \\
142.5 & \ion{Fe}{11} & 2.4 & 10.4 & 1.0 & \phs \phn \phn 5.1 \\
142.5 & \ion{Fe}{12} & 2.0 & \phn 8.4 & 0.9 & \phs \phn \phn 6.6 \\
142.5 & \ion{Fe}{13} & 1.6 & \phn 8.2 & 1.2 & \phs \phn \phn 15.0 \\
142.5 & \ion{Fe}{14} & 2.0 & \phn 8.7 & 1.9 & \phs \phn \phn 23.9 \\
142.5 & \ion{Fe}{15} & 2.2 & \phn 8.9 & 2.1 & \phs \phn \phn 23.3 \\
190.0 & \ion{Fe}{11} & 0.3 & \phn 8.8 & 4.7 & $-411$ \\
190.0 & \ion{Fe}{12} & 0.8 & \phn 7.5 & 3.5 & \phs \phn \phn 36.8 \\
190.0 & \ion{Fe}{13} & 1.2 & \phn 7.3 & 3.1  &  \phs \phn \phn 36.3 \\
190.0 & \ion{Fe}{14} & 1.5 & \phn 7.3 & 3.2 &  \phs \phn \phn 43.6 \\
190.0 & \ion{Fe}{15} & 2.2 & \phn 7.5 & 3.4 & \phs \phn \phn 12.8 \\
\enddata
\end{deluxetable}

Examining the amplitude of the oscillations in each time interval
as a function of the temperature of line formation shows no clear
trend. The data set starting at 113.9 minutes seems to show
evidence for a decrease in amplitude with increasing temperature,
while the data set beginning at 190.0 minutes shows the opposite
trend. Similarly, there is no clear trend in the periods of the
oscillations in each data set as a function of temperature.

\subsection{Intensity Oscillations}

\begin{figure}[b]
\plotone{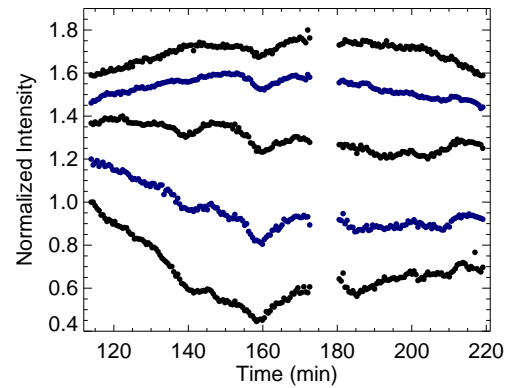} 
\caption{Normalized intensity data averaged over the 16 detector
  rows from $-200$\arcsec\ to $-215$\arcsec. The emission lines
  shown from top to bottom are \ion{Fe}{11} 188.23~\AA,
  \ion{Fe}{12} 195.12~\AA, \ion{Fe}{13} 202.04~\AA, \ion{Fe}{14}
  274.20~\AA, and \ion{Fe}{15} 284.16~\AA. Time is measured in
  minutes from the start of the sit-and-stare data at 18:13:06
  UT. Data for each emission line are displaced from the next
  set by 0.2.}
\label{fig:intensity_average}
\end{figure}

If the observed Doppler shift oscillations are acoustic in
nature, then they should also be visible in the intensity data.
For a linear sound wave $v = c_{\textrm{s}} \delta \rho/\rho$,
where $v$ is the amplitude of the wave, $c_{\textrm{s}}$ is the
sound speed, and $\delta \rho$ is the density perturbation on the
background density $\rho$. Taking an amplitude of 2 km s$^{-1}$,
yields values of $\delta \rho/\rho$ of around 1\%. Since the
intensity fluctuation, $\delta I/I$, is proportional to $2 \delta
\rho/\rho$, we expect only about a 2\% fluctuation in the
measured intensity. This number could of course increase if the
actual velocity is much larger due to a large difference between
the line-of-sight and the direction of the coronal structure
being measured. Figure~\ref{fig:intensity_average} shows the
measured intensity data averaged over the same locations as the
Doppler shift data shown in Figure~\ref{fig:doppler_average}. The
data show little or no evidence for oscillations with the periods
measured in the Doppler shift data. This is borne out by a
periodogram analysis of the time series in the figure, which show
no significant peaks.

\begin{figure}
\plotone{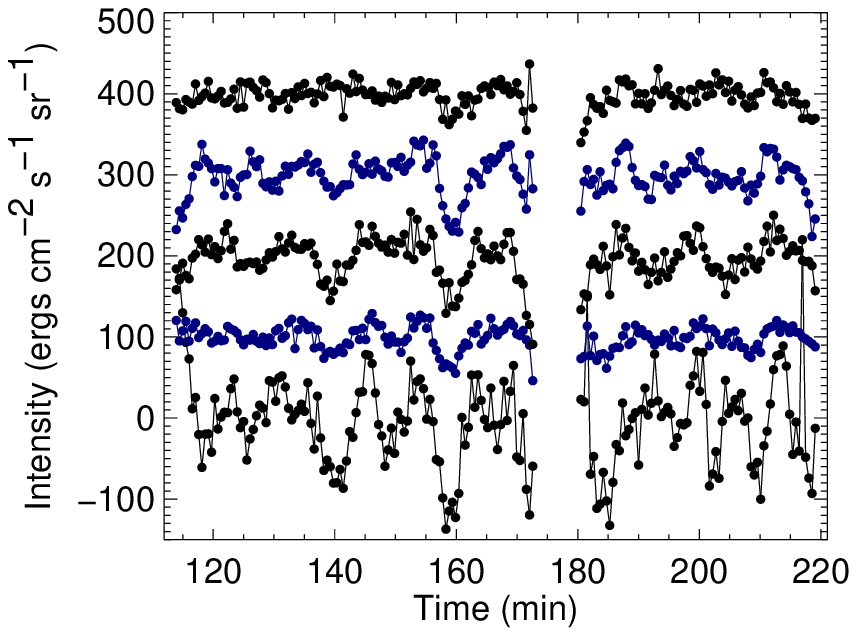} 
\caption{Detrended intensity data averaged over the 16 detector
  rows from $-200$\arcsec\ to $-215$\arcsec. The emission lines
  shown from top to bottom are \ion{Fe}{11} 188.23~\AA,
  \ion{Fe}{12} 195.12~\AA, \ion{Fe}{13} 202.04~\AA, \ion{Fe}{14}
  274.20~\AA, and \ion{Fe}{15} 284.16~\AA. Time is measured in
  minutes from the start of the sit-and-stare data at 18:13:06
  UT. Data for each emission line are displaced from the next
  set by 100 ergs cm$^{-2}$ s$^{-1}$ sr$^{-1}$.}
\label{fig:detrend_i}
\end{figure}

If, however, we detrend the data by subtracting the gradually
evolving background signal, there is some evidence for an
oscillatory signal. Figure~\ref{fig:detrend_i} shows the data in
Figure~\ref{fig:intensity_average} with a background consisting
of a 10-minute average of the data centered on each data point
subtracted. All the emission lines show some evidence for an
oscillatory signal, with the \ion{Fe}{13} 202.04~\AA\ emission
line being the most obvious.

\begin{figure}
\plotone{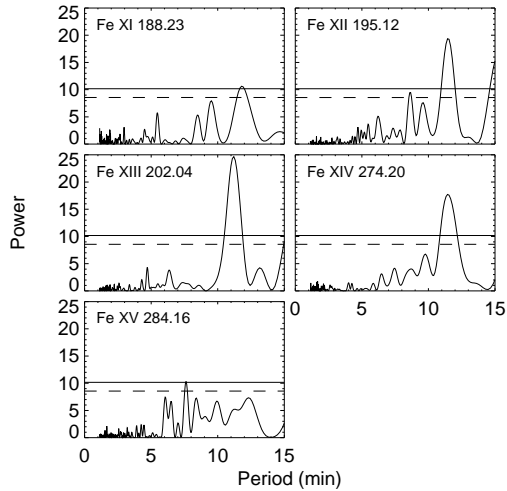}  
\caption{Periodograms computed for the detrended intensity data
  shown in normalized form in Figure~\ref{fig:intensity_average}.
  The solid horizontal line on each plot indicates the power
  equivalent to a calculated false alarm probability of 1\% and
  the dashed line is for a false alarm probability of 5\%.}
\label{fig:pgram_i}
\end{figure}

\begin{figure}
\plotone{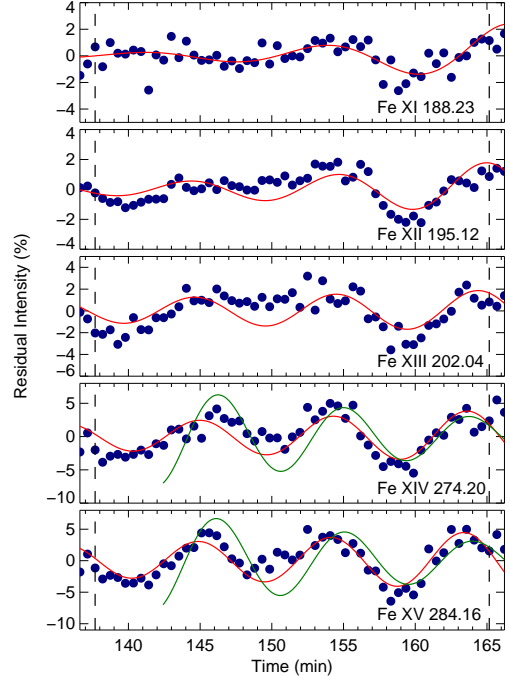} 
\caption{Decaying sine wave fits to the detrended intensity data
  beginning roughly 138 minutes after the start of the
  sit-and-stare observation. The intensities have been converted
  to residual intensities by taking the difference between the
  intensity at each data point and the running mean intensity and
  dividing it by the running mean. The plotted data has also had
  the polynomial background used in the fit removed. Vertical
  dashed lines show the time range used in the fitting. Also
  plotted as green curves on the panels for the \ion{Fe}{14} and
  \ion{Fe}{15} intensity data are the fitting results for the
  \ion{Fe}{14} and \ion{Fe}{15} Doppler shift data. Those plots
  show $-v(t)$.}
\label{fig:fit_fig_i}
\end{figure}

Figure~\ref{fig:pgram_i} shows periodograms constructed for the
emission lines shown in Figure~\ref{fig:detrend_i}. Each
periodogram shows a significant peak. The periods for the
strongest peak in each periodogram are listed in
Table~\ref{table:doppler_periods}. The periods are generally
consistent with those determined from the Doppler shift data.
Also listed in the table is an estimate of the intensity
fluctuation in each emission line. This was obtained by computing
the standard deviation of the detrended intensity in the time
series for each emission line and then dividing the result by the
average intensity. The values are roughly consistent with those
expected based on the $\delta v$ estimates listed in
Table~\ref{table:doppler_periods}.

While the oscillatory signal is much less strong in the detrended
intensity data than in the Doppler shift data, it is possible to
fit some of the data in roughly the same time intervals that were
used for the results listed in
Table~\ref{table:oscillation_d_fits}.
Table~\ref{table:oscillation_i_fits} shows the resulting fit
information. Note that intensity oscillation fits were not
possible for all the lines for which the Doppler shift data could
be fitted. Figure~\ref{fig:fit_fig_i} shows an example of the
fits to the detrended intensity data. For these plots the
intensity has been converted to a residual intensity expressed in
\% by taking the difference between each data point and the
10-minute average and dividing it by the 10-minute average. The
plotted points have also had the polynomial background
subtracted.To facilitate comparisons with the Doppler shift data
we have also plotted as green curves on the panels for the
\ion{Fe}{14} and \ion{Fe}{15} intensity data the fitting results
for the \ion{Fe}{14} and \ion{Fe}{15} Doppler shift data.

Comparison of the fits in Figure~\ref{fig:fit_fig_i} with those
in Figure~\ref{fig:fit_fig_v} shows some similarities and many
differences between the two data sets. For the lines with a
reasonably strong signal, the periods measured for the residual
intensity data are similar to those measured for the Doppler
shifts. In addition, the residual intensities show the same trend
to larger amplitudes as the temperature of line formation
increases. The intensity oscillations clearly start earlier than
the Doppler shift oscillations. Moreover, while the Doppler shift
oscillations are damped, the intensity oscillations appear to
grow with time over the same interval. This appears to be the
case for the other time intervals as well. Fitting the detrended
intensity signal is more challenging than fitting the Doppler
shift signal. Also, implicit in the fitting is the idea that a
damped sine wave can fully represent what is probably a more
complex signal. Thus, we are reluctant to read too much into this
growth in the intensity until we can determine from additional
data sets if it is a common phenomenon.

\begin{deluxetable}{llcccc}
\tablecaption{Intensity Oscillation
Properties\label{table:oscillation_i_fits}} \tablewidth{0pt}
\tablehead{
   \colhead{$t_0$} & \colhead{} &
  \colhead{$A_0$} & \colhead{$P$} & \colhead{$\phi$} &
  \colhead{$\lambda^{-1}$} \\
  \colhead{(min)} & \colhead{Ion} & 
  \colhead{(\%)} &
  \colhead{(min)} & \colhead{(rad)} & \colhead{(min)} 
} \startdata
113.9 & \ion{Fe}{12} & 2.1 & \phn 8.0 & 4.2 & \phs \phn 8.2 \\
113.9 & \ion{Fe}{13} & 1.0 & 12.3 & 4.4 &  \phs 37.6 \\
137.7 & \ion{Fe}{11} & 0.2 & 12.7 & 0.0 & $-11.4$ \\
137.7 & \ion{Fe}{12} & 0.4 & 10.3 & 3.9 & $-17.8$ \\
137.7 & \ion{Fe}{13} & 1.1 & \phn 9.9 & 3.5 & $-50.2$ \\
137.7 & \ion{Fe}{14} & 2.1 & \phn 9.3 & 3.0 & $-41.8$ \\
137.7 & \ion{Fe}{15} & 2.6 & \phn 9.3 & 3.0 & $-48.6$ \\
190.0 & \ion{Fe}{11} & 0.5 & \phn 8.8 &  5.0 & $-81.1$ \\
190.0 & \ion{Fe}{12} & 0.4 & 13.7 & 3.7 & $-32.6$ \\
190.0 & \ion{Fe}{13} & 1.1 & 13.5 & 3.8 &  $-83.6$ \\
190.0 & \ion{Fe}{14} & 0.8 & 12.8 & 2.5 &  $-20.3$ \\
190.0 & \ion{Fe}{15} & 1.3 & \phn 6.9 & 5.6 & $-22.8$ \\
\enddata
\end{deluxetable}

An important factor in determining the nature of the oscillations
is the phase difference between the Doppler shift signal and the
intensity signal. Comparing the phases of the fits listed in
Table~\ref{table:oscillation_d_fits} with those in
Table~\ref{table:oscillation_i_fits} is difficult because the
periods are not identical, but, since the periods are close, the
differences do not significantly alter any conclusions that we
might draw. To facilitate this comparison we have plotted as
green curves on the panels for the \ion{Fe}{14} and \ion{Fe}{15}
intensity data in Figure~\ref{fig:fit_fig_i} the fitting results
for the \ion{Fe}{14} and \ion{Fe}{15} Doppler shift data. For the
Doppler shift data, we plot $-v(t)$. The curves show that for
these two ions, the intensity variations are close to $180\degr$
out of phase with the Doppler shift variations. Since we define
the Doppler shift as $c\, \delta \lambda/\!\lambda$, this means
that the peak intensity corresponds to a blueshift, indicating an
upward propagating wave. For the other two time intervals, the
situation is more ambiguous. Examination of the tables shows that
in many cases, the periods are significantly different for the
same Doppler shift and intensity data in the same line. In those
cases where the periods are close (e.g., \ion{Fe}{12} at
$t_0=113.9$ minutes and \ion{Fe}{15} at $t_0=190.0$ minutes),
examination of the plots similar to Figures~\ref{fig:fit_fig_v}
and \ref{fig:fit_fig_i} shows the same $180\degr$ phase shift,
again indicating upwardly propagating oscillations.

Even for the cases where the periods are close, the agreements in
the phases are only approximate. For the \ion{Fe}{14} and
\ion{Fe}{15} intensity and Doppler shift fits shown in
Figure~\ref{fig:fit_fig_i}, the intensity oscillation leads the
Doppler shift oscillation by a small fraction of a period. For
both the \ion{Fe}{12} data at $t_0=113.9$ minutes and the
\ion{Fe}{15} data at $t_0=190.0$ minutes, the Doppler shift
oscillation leads the intensity oscillation by a fraction of a
period. In both cases, this difference is less than the $1/4$
period expected for a standing-mode MHD wave \citep{Sakurai2002}.
\citet{Wang2009} observed propagating waves with periods in the
four to six minute range in EIS active region observations. They
noted that for most cases the Doppler shift and intensity
oscillations were nearly in phase. In the cases where there was a
difference, the phase of the intensity was earlier than the
Doppler shift, as is the case for the data shown in
Figures~\ref{fig:fit_fig_v} and \ref{fig:fit_fig_i}. Theoretical
modeling of propagating slow waves with periods near five minutes
show that thermal conduction can produce phase shifts between the
intensity and Doppler shifts \citep{Owen2009}. Further study of
EIS data sets where both the Doppler shift and intensity can be
fitted could provide valuable constraints on these models.

\subsection{Density Oscillations}

The electron density is one factor in determining the Alfv\'{e}n
speed in the oscillating plasma. Moreover, for magnetoacoustic
fluctuations, we expect the density to also oscillate. Thus a
direct measurement can aid in disentangling the nature of the
oscillations. The sit-and-stare observations included
density-sensitive line pairs of \ion{Fe}{12}
(186.88~\AA/195.12~\AA) and \ion{Fe}{13} (203.83~\AA/202.04~\AA).
Using data from version 5.2 of the CHIANTI database
\citep{Landi2006,Dere1997}, we computed the electron density at
each time for the row-averaged data. For \ion{Fe}{12}, CHIANTI
uses energy levels and radiative decay rates from
\citet{DelZanna2005}, electron collision strengths from
\citet{Storey2005}, and proton collision rate coefficients from
\citet{Landman1978}. For \ion{Fe}{13}, CHIANTI uses energy levels
from \citet{Penn1994}, \citet{Jupen1993}, and version 1.0 of the
NIST database; radiative decay rates from \citet{Young2004},
electron collision strengths from \citet{Gupta1998}, and proton
collision rate coefficients from \citet{Landman1975}. These
diagnostics are discussed in detail in \citet{Young2009}.

\begin{figure}
\plotone{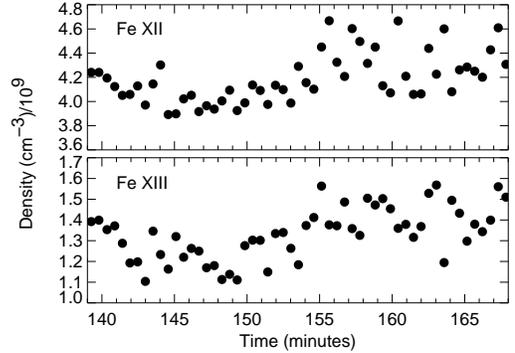} 
\caption{Electron density determined using the \ion{Fe}{12}
  186.88~\AA/195.12~\AA\ ratio (top) and the \ion{Fe}{13}
  203.83~\AA/202.04~\AA\ ratio (bottom) for the data beginning
  roughly 138 minutes after the start of the sit-and-stare
  observation.}
\label{fig:density}
\end{figure}

Figure~\ref{fig:density} shows the derived electron densities as
a function of time for the same time interval shown in
Figures~\ref{fig:fit_fig_v} and \ref{fig:fit_fig_i}. Both sets of
derived densities show the same overall time behavior, but the
absolute values differ by nearly a factor of three. Differences
between these diagnostics have been noted before
\citep[e.g.,][]{Young2009}, and are thought to be due to issues
with the atomic data. It is not yet clear which of the values
should be considered the most reliable.

\begin{figure}
\plotone{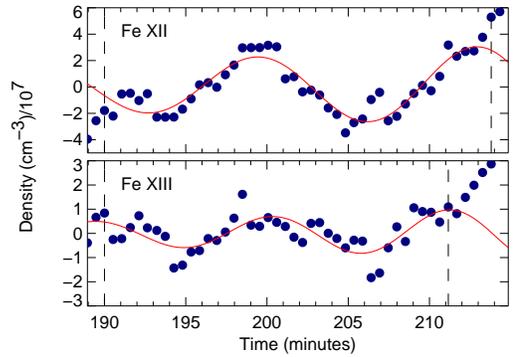} 
\caption{Decaying sine wave fits to the smoothed electron density
  determined using the \ion{Fe}{12} 186.88~\AA/195.12~\AA\ ratio
  (top) and the \ion{Fe}{13} 203.83~\AA/202.04~\AA\ ratio
  (bottom) for the data beginning 190 minutes after the start of
  the sit-and-stare observation along with. The plotted data has
  had the polynomial background removed. Vertical dashed lines
  show the time range used in the fitting.}
\label{fig:density_det_fit}
\end{figure}

Neither of the density time series shown in the figure display
any evidence for the oscillations detected in the detrended
intensity data. If we smooth the density time series with a
10-minute running mean, there is some evidence for oscillatory
behavior over the time range beginning at 190 minutes. Decaying
sine wave fits to that region are show in
Figure~\ref{fig:density_det_fit}. For the \ion{Fe}{12} time
series the fit has an amplitude of $1.9 \times 10^7$~cm$^{-3}$, a
period of 13.5 minutes, a decay time of $-46.4$ minutes, and a
phase of 3.5 radians, all consistent with the values listed for
the \ion{Fe}{12} data in Table~\ref{table:oscillation_i_fits} for
this time interval. For the \ion{Fe}{13 } time series the fit has
an amplitude of $5.1 \times 10^6$~cm$^{-3}$, a period of 10.9
minutes, a decay time of $-25.1$ minutes, and a phase of 1.9
radians. With the exception of the phase, these values are
generally consistent with the \ion{Fe}{13} data in
Table~\ref{table:oscillation_i_fits} for this time interval. The
amplitudes for the \ion{Fe}{12} and \ion{Fe}{13} fits are 0.5\%
and 0.4\% of the average density in the time interval. Since the
observed intensity fluctuations are only about 1\%, though, and
we expect $\delta I/I$ to be proportional to $2 \delta
\rho/\rho$, they are consistent with the observed intensity
oscillations.

\subsection{Underlying Chromospheric Behavior}

As we pointed out earlier, SOT obtained a time sequence of
\ion{Ca}{2} H images that is co-temporal with the EIS
sit-and-stare observations starting at 19:50 UT and ending at
21:40 UT, with a constant cadence of 60~s. To further use this
data, we applied the standard reduction procedure provided by the
SOT team that is available in SolarSoft. The images of the
\ion{Ca}{2} H sequence were then carefully aligned using Fourier
cross-correlation techniques to remove residual jitter and the
drift of the SOT correlation tracker.

As can be seen in the lower left panel of
Figure~\ref{fig:coalign}, the EIS slit covers the network bright
point, which is the focus of the chromospheric analysis.
Considering the accuracy of the coalignment and the spatial
averaging applied to the EIS data, we also average the
\ion{Ca}{2} H signal.

\begin{figure}
\plotone{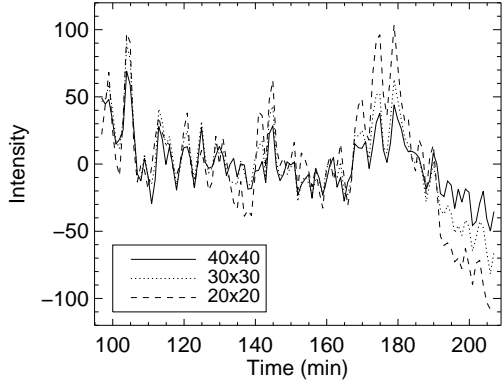} 
\caption{\ion{Ca}{2} H intensity data averaged over different
  spatial areas centered on the magnetic feature highlighted in
  Figure~\ref{fig:coalign}. Each data set has had the mean value
  subtracted. Spatial areas are given in SOT pixels, which are
  0.10896\arcsec in size, giving areas of $2.18\arcsec \times
  2.18\arcsec$, $3.77\arcsec \times 3.77\arcsec$, and
  $4.36\arcsec \times 4.36\arcsec$, respectively.}
\label{fig:ca_i}
\end{figure}

\begin{figure}
\plotone{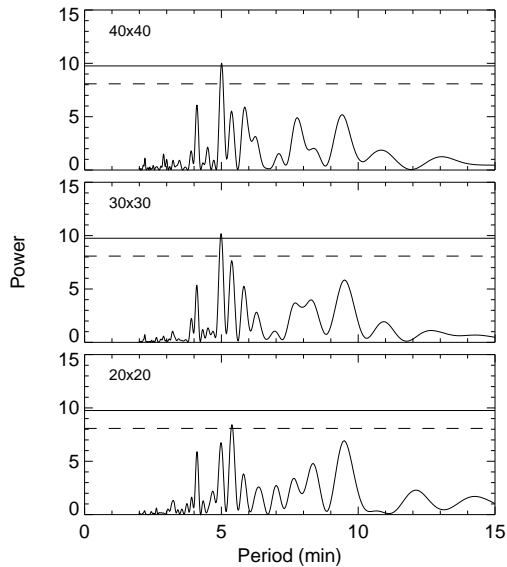} 
\caption{Periodograms computed for the detrended \ion{Ca}{2} H
  intensity data. The size of the area averaged over in pixels in
  indicated in each plot. The solid horizontal line on each plot
  indicates the power equivalent to a calculated false alarm
  probability of 1\% and the dashed line is for a false alarm
  probability of 5\%. Spatial areas are given in SOT pixels,
  which are 0.10896\arcsec in size, giving areas of $2.18\arcsec
  \times 2.18\arcsec$, $3.77\arcsec \times 3.77\arcsec$, and
  $4.36\arcsec \times 4.36\arcsec$, respectively.}
\label{fig:ca_periodograms}
\end{figure}

Figure~\ref{fig:ca_i} shows the time history of the \ion{Ca}{2}
H-line intensity for three different sized spatial areas centered
on the feature shown in Figure~\ref{fig:coalign}. The sizes of
the regions are given in SOT pixels, which are $0.10896$\arcsec\
in size. As expected for chromospheric lines, all three averaged
data sets show evidence for intensity oscillations with a period
near 5 minutes. The data, however, are quite noisy and
periodograms constructed from them show no significant peaks.
Detrending the data, however, results in periodograms with
significant peaks. Figure~\ref{fig:ca_periodograms} shows
periodograms for the three data sets with each detrended by
subtracting a 9-minute running mean from each data point. The
periodograms show clear evidence for oscillations near 5 minutes.

There is also some evidence for power at periods between 9 and 10
minutes, but with more than a 5\% false alarm probability. If we
instead detrend the data with an 11-minute running mean, then the
peak between 9 and 10 minutes becomes more prominent and is at or
above the 5\% false alarm probability level for the
$30~\mathrm{pixel} \times 30~\mathrm{pixel}$ and
$20~\mathrm{pixel} \times 20~\mathrm{pixel}$ data sets. Wavelet
analysis of the data sets detrended with both a 9- and 11-minute
running mean shows significant signal near 9 minutes.

Note that the data in Figure~\ref{fig:ca_i} shows three larger
peaks in all three data sets. The first peak, near 110 minutes,
occurs before the beginning of the EIS data plots in
Figures~\ref{fig:doppler_average} and \ref{fig:detrend_i}. The
other two, at roughly 140 and 170 minutes, come just before
significant oscillatory signal is observed in the EIS Doppler
shift and detrended intensity data. It is tempting to suggest
that these enhancements correspond to chromospheric events that
resulted in the oscillations observed with EIS.

In principle the EIS \ion{He}{2} 256.32~\AA\ data can bridge the
gap in temperature between the SOT \ion{Ca}{2} data and the Fe
lines formed at higher temperatures that are the main focus of
this study. In practice the \ion{He}{2} data are challenging to
analyze. The line is closely blended with a \ion{Si}{10} line at
256.37~\AA\ along with a smaller contribution from \ion{Fe}{10}
and \ion{Fe}{13} ions at slightly longer wavelengths
\citep{Brown2008}. In an effort to see if a connection can be
made, we have made two-component Gaussian fits to the
row-averaged \ion{He}{2} data. Periodograms of the resulting
Doppler shift data show no significant periods. Periodograms of
the \ion{He}{2} fitted intensities detrended with a 10-minute
running mean show no peaks at the 1\% false alarm probably level
and one peak with a period between 12 and 13 minutes at the 5\%
false alarm probability level. Examining the detrended
\ion{He}{2} intensity data, we do not see a peak in the data at
140 minutes, but do see a significant increase at 170 minutes.
Thus the \ion{He}{2} data only weakly support our suggestion that
the enhancements seen in the \ion{Ca}{2} data correspond to
chromospheric events that result in the oscillations observed
with EIS in the Fe lines.

\section{DISCUSSION AND CONCLUSIONS}

As we pointed out in \S\ref{intro}, a number of investigations
have detected Doppler shift oscillations with EIS. Based on the
phase differences between the Doppler shift and the intensity, we
believe that the signals we have detected are upwardly
propagating magnetoacoustic waves. The periods we detect are
between 7 and 14 minutes. For a set of observations that begins
at a particular time, there is considerable scatter in the
measured periods and amplitudes. This is probably due to the
relatively weak signal we are analyzing. But it may also be an
indication that a simple sine wave fit is not a good
representation of the data. It is likely that each line-of-sight
passes through a complex, time-dependent dynamical system. While
a single flux tube may respond to an oscillatory signal by
exhibiting a damped sine wave in the Doppler shift, a more
complex line-of-sight may display a superposition of waves.

Coalignment of the EIS data with both SOT and MDI magnetograms
shows that the portion of the EIS slit analyzed in this study
corresponds to a unipolar flux concentration. SOT \ion{Ca}{2}
images show that the intensity of this feature exhibits 5-minute
oscillations typical of chromospheric plasma, but also exhibits
some evidence for longer period oscillations in the time range
detected by EIS. Moreover, the \ion{Ca}{2} intensity data show
that the oscillations observed in EIS are related to significant
enhancements in the \ion{Ca}{2} intensities, suggesting that a
small chromospheric heating event triggered the observed EIS
response.

\citet{Wang2009} also detected propagating slow magnetoacoustic
waves in an active region observed with EIS, which they
associated with the footpoint of a coronal loop. While the
oscillation periods they measured---5 minutes---were smaller than
those detected here, many of the overall characteristics we see
are the same. In each case, the oscillation only persists for a
few cycles and the phase relationship indicates an upwardly
propagating wave. In contrast with their results, however, we do
not see a consistent trend for the oscillation amplitude to
decrease with increasing temperature of line formation.
Examination of both the Doppler shift data in
Figure~\ref{fig:doppler_average} and the results in
Table~\ref{table:oscillation_d_fits} shows that in one case the
amplitude has a tendency to decrease with increasing temperature
of line formation (oscillation beginning at 113.9 minutes) and in
another case the amplitude clearly increases with increasing
temperature of line formation (oscillation beginning at 190
minutes). Thus it does not appear that the results reported by
\citet{Wang2009} are always the case. \citet{O'Shea2002} noted
that for oscillations observed above a sunspot the amplitude
decreased with increasing temperature until the temperature of
formation of \ion{Mg}{10}, which is formed at roughly 1~MK. They
then saw an increase in amplitude in emission from \ion{Fe}{16}.
All the EIS lines we have included in this study have
temperatures of formation greater than 1~MK.

Combined EIS and SUMER polar coronal hole observations have also
shown evidence propagating slow magnetoacoustic waves
\citep{Banerjee2009}. These waves have periods in the 10--30
minute range. These waves appear to be more like those we observe
in that they are have periods longer than those studied by
\citet{Wang2009}.

\citet{Wang2009} suggested that the waves they observed were the
result of leakage of photospheric p-mode oscillations upward into
the corona. The longer periods we and \citet{Banerjee2009}
observe are probably not related to p-modes. Instead, we
speculate that the periods of the waves are related to the
impulsive heating which may be producing them. If an instability
is near the point where rather than generating a catastrophic
release of energy it wanders back and forth between generating
heating and turning back off, waves would be created. The heating
source could be at a single location, or, for example, locations
near each other where instability in one place causes a second
nearby location to go unstable and begin heating plasma. In this
view, the periods provide some insight into the timescale for the
heating to rise and fall and thus may be able to place limits on
possible heating mechanisms.

The behavior of slow magnetoacoustic oscillations as a function
of temperature has been the subject of considerable theoretical
work. It is generally believed that the damping of the waves is
due to thermal conduction
\citep[e.g.,][]{DeMoortel2003,Klimchuk2004}. Because thermal
conduction scales as a high power of the temperature, conductive
damping should be stronger for oscillations detected in higher
temperature emission lines \citep[e.g.,][]{Porter1994,Ofman2002}.
Earlier EIS observations of the damping of standing slow
magnetoacoustic waves, however, show that this is not always the
case \citep{Mariska2008}. Thus, the temperature behavior of both
the amplitude of the oscillations and the damping, differ from
some earlier results. We believe that additional observations
will be required to understand fully the physical picture of what
is occurring in the low corona when oscillations are observed.
Given the complex set of structures that may be in the line of
sight to any given solar location under the EIS slit, we are not
entirely surprised that different data sets should yield
different results, which in some cases differ from models. For
example, none of the current models for oscillations in the outer
layers of the solar atmosphere take into account the possibility
that what appear to be single structures in the data might
actually be bundles of threads with differing physical
conditions.

Our observations along with others
\citep[e.g.,][]{Wang2009,Wang2009a,Banerjee2009} show that
low-amplitude upwardly propagating slow magnetoacoustic waves are
not uncommon in the low corona. The periods observed to date
range from 5 minutes to 30 minutes. In all cases, however, the
wave amplitudes are too small to contribute significantly to
coronal heating. But understanding how the waves are generated
and behave as a function of line formation temperature and the
structure of the magnetic field should lead to a more complete
understanding of the structure of the low corona and its
connection with the underlying portions of the atmosphere.
Instruments like those on \textit{Hinode} that can simultaneously
observe both the chromosphere and the corona, should provide
valuable additional insight into these waves as the new solar
cycle rises and more active regions become available for study.

\acknowledgments \textit{Hinode} is a Japanese mission developed,
launched, and operated by ISAS/JAXA in partnership with NAOJ,
NASA, and STFC (UK). Additional operational support is provided
by ESA and NSC (Norway). The authors acknowledge support from the
NASA \textit{Hinode} program. CHIANTI is a collaborative project
involving NRL (USA), RAL (UK), MSSL (UK), the Universities of
Florence (Italy) and Cambridge (UK), and George Mason University
(USA). We thank the anonymous referee for his or her very helpful
comments.

\bibliographystyle{apj}
\bibliography{allrefs}

\begin{thebibliography}{39}
\expandafter\ifx\csname natexlab\endcsname\relax\def\natexlab#1{#1}\fi

\bibitem[{{Aschwanden} {et~al.}(1999){Aschwanden}, {Fletcher}, {Schrijver}, \&
  {Alexander}}]{Aschwanden1999}
{Aschwanden}, M.~J., {Fletcher}, L., {Schrijver}, C.~J., \& {Alexander}, D.
  1999, \apj, 520, 880

\bibitem[{{Banerjee} {et~al.}(2009){Banerjee}, {Teriaca}, {Gupta}, {Imada},
  {Stenborg}, \& {Solanki}}]{Banerjee2009}
{Banerjee}, D., {Teriaca}, L., {Gupta}, G.~R., {Imada}, S., {Stenborg}, G., \&
  {Solanki}, S.~K. 2009, \aap, 499, L29

\bibitem[{{Berghmans} \& {Clette}(1999)}]{Berghmans1999}
{Berghmans}, D., \& {Clette}, F. 1999, \solphys, 186, 207

\bibitem[{{Bevington}(1969)}]{Bevington1969}
{Bevington}, P.~R. 1969, {Data reduction and error analysis for the physical
  sciences} (New York: McGraw-Hill)

\bibitem[{{Brown} {et~al.}(2008){Brown}, {Feldman}, {Seely}, {Korendyke}, \&
  {Hara}}]{Brown2008}
{Brown}, C.~M., {Feldman}, U., {Seely}, J.~F., {Korendyke}, C.~M., \& {Hara},
  H. 2008, \apjs, 176, 511

\bibitem[{{Culhane} {et~al.}(2007)}]{Culhane2007}
{Culhane}, J.~L., {et~al.} 2007, \solphys, 243, 19

\bibitem[{{De Moortel} \& {Hood}(2003)}]{DeMoortel2003}
{De Moortel}, I., \& {Hood}, A.~W. 2003, \aap, 408, 755

\bibitem[{{DeForest} \& {Gurman}(1998)}]{DeForest1998}
{DeForest}, C.~E., \& {Gurman}, J.~B. 1998, \apjl, 501, L217

\bibitem[{{Del Zanna} \& {Mason}(2005)}]{DelZanna2005}
{Del Zanna}, G., \& {Mason}, H.~E. 2005, \aap, 433, 731

\bibitem[{{Dere} {et~al.}(1997){Dere}, {Landi}, {Mason}, {Monsignori Fossi}, \&
  {Young}}]{Dere1997}
{Dere}, K.~P., {Landi}, E., {Mason}, H.~E., {Monsignori Fossi}, B.~C., \&
  {Young}, P.~R. 1997, \aaps, 125, 149

\bibitem[{{Gupta} \& {Tayal}(1998)}]{Gupta1998}
{Gupta}, G.~P., \& {Tayal}, S.~S. 1998, \apj, 506, 464

\bibitem[{{Horne} \& {Baliunas}(1986)}]{Horne1986}
{Horne}, J.~H., \& {Baliunas}, S.~L. 1986, \apj, 302, 757

\bibitem[{{Jupen} {et~al.}(1993){Jupen}, {Isler}, \& {Trabert}}]{Jupen1993}
{Jupen}, C., {Isler}, R.~C., \& {Trabert}, E. 1993, \mnras, 264, 627

\bibitem[{{Klimchuk} {et~al.}(2004){Klimchuk}, {Tanner}, \& {De
  Moortel}}]{Klimchuk2004}
{Klimchuk}, J.~A., {Tanner}, S.~E.~M., \& {De Moortel}, I. 2004, \apj, 616,
  1232

\bibitem[{{Kosugi} {et~al.}(2007)}]{Kosugi2007}
{Kosugi}, T., {et~al.} 2007, \solphys, 243, 3

\bibitem[{{Landi} {et~al.}(2006){Landi}, {Del Zanna}, {Young}, {Dere}, {Mason},
  \& {Landini}}]{Landi2006}
{Landi}, E., {Del Zanna}, G., {Young}, P.~R., {Dere}, K.~P., {Mason}, H.~E., \&
  {Landini}, M. 2006, \apjs, 162, 261

\bibitem[{{Landman}(1975)}]{Landman1975}
{Landman}, D.~A. 1975, \aap, 43, 285

\bibitem[{{Landman}(1978)}]{Landman1978}
---. 1978, \apj, 220, 366

\bibitem[{{Mariska} {et~al.}(2008){Mariska}, {Warren}, {Williams}, \&
  {Watanabe}}]{Mariska2008}
{Mariska}, J.~T., {Warren}, H.~P., {Williams}, D.~R., \& {Watanabe}, T. 2008,
  \apjl, 681, L41

\bibitem[{{Ofman} \& {Wang}(2002)}]{Ofman2002}
{Ofman}, L., \& {Wang}, T. 2002, \apjl, 580, L85

\bibitem[{{O'Shea} {et~al.}(2002){O'Shea}, {Muglach}, \& {Fleck}}]{O'Shea2002}
{O'Shea}, E., {Muglach}, K., \& {Fleck}, B. 2002, \aap, 387, 642

\bibitem[{{Owen} {et~al.}(2009){Owen}, {De Moortel}, \& {Hood}}]{Owen2009}
{Owen}, N.~R., {De Moortel}, I., \& {Hood}, A.~W. 2009, \aap, 494, 339

\bibitem[{{Penn} \& {Kuhn}(1994)}]{Penn1994}
{Penn}, M.~J., \& {Kuhn}, J.~R. 1994, \apj, 434, 807

\bibitem[{{Porter} {et~al.}(1994){Porter}, {Klimchuk}, \&
  {Sturrock}}]{Porter1994}
{Porter}, L.~J., {Klimchuk}, J.~A., \& {Sturrock}, P.~A. 1994, \apj, 435, 482

\bibitem[{{Press} \& {Rybicki}(1989)}]{Press1989}
{Press}, W.~H., \& {Rybicki}, G.~B. 1989, \apj, 338, 277

\bibitem[{{Roberts}(2000)}]{Roberts2000}
{Roberts}, B. 2000, \solphys, 193, 139

\bibitem[{{Roberts} {et~al.}(1983){Roberts}, {Edwin}, \& {Benz}}]{Roberts1983}
{Roberts}, B., {Edwin}, P.~M., \& {Benz}, A.~O. 1983, \nat, 305, 688

\bibitem[{{Roberts} {et~al.}(1984){Roberts}, {Edwin}, \& {Benz}}]{Roberts1984}
---. 1984, \apj, 279, 857

\bibitem[{{Roberts} \& {Nakariakov}(2003)}]{Roberts2003}
{Roberts}, B., \& {Nakariakov}, V.~M. 2003, in NATO Science Series: II:
  Mathematics, Physics and Chemistry, Vol. 124, Turbulence, Waves and
  Instabilities in the Solar Plasma, ed. R.~Erdelyi, K.~Petrovay, B.~Roberts,
  \& M.~J. Aschwanden (Kluwer Academic Publishers, Dordrecht), 167--192

\bibitem[{{Rosenberg}(1970)}]{Rosenberg1970}
{Rosenberg}, H. 1970, \aap, 9, 159

\bibitem[{{Sakurai} {et~al.}(2002){Sakurai}, {Ichimoto}, {Raju}, \&
  {Singh}}]{Sakurai2002}
{Sakurai}, T., {Ichimoto}, K., {Raju}, K.~P., \& {Singh}, J. 2002, \solphys,
  209, 265

\bibitem[{{Storey} {et~al.}(2005){Storey}, {Del Zanna}, {Mason}, \&
  {Zeippen}}]{Storey2005}
{Storey}, P.~J., {Del Zanna}, G., {Mason}, H.~E., \& {Zeippen}, C.~J. 2005,
  \aap, 433, 717

\bibitem[{{Van Doorsselaere} {et~al.}(2008){Van Doorsselaere}, {Nakariakov},
  {Young}, \& {Verwichte}}]{VanDoorsselaere2008}
{Van Doorsselaere}, T., {Nakariakov}, V.~M., {Young}, P.~R., \& {Verwichte}, E.
  2008, \aap, 487, L17

\bibitem[{{Wang} {et~al.}(2002){Wang}, {Solanki}, {Curdt}, {Innes}, \&
  {Dammasch}}]{Wang2002}
{Wang}, T., {Solanki}, S.~K., {Curdt}, W., {Innes}, D.~E., \& {Dammasch}, I.~E.
  2002, \apjl, 574, L101

\bibitem[{{Wang} {et~al.}(2009{\natexlab{a}}){Wang}, {Ofman}, \&
  {Davila}}]{Wang2009}
{Wang}, T.~J., {Ofman}, L., \& {Davila}, J.~M. 2009{\natexlab{a}}, \apj, 696,
  1448

\bibitem[{{Wang} {et~al.}(2009{\natexlab{b}}){Wang}, {Ofman}, {Davila}, \&
  {Mariska}}]{Wang2009a}
{Wang}, T.~J., {Ofman}, L., {Davila}, J.~M., \& {Mariska}, J.~T.
  2009{\natexlab{b}}, \aap, 503, L25

\bibitem[{{Wuelser} {et~al.}(2004)}]{Wuelser2004}
{Wuelser}, J.-P., {et~al.} 2004, in Society of Photo-Optical Instrumentation
  Engineers (SPIE) Conference Series, ed. S.~{Fineschi} \& M.~A. {Gummin}, Vol.
  5171, 111

\bibitem[{{Young}(2004)}]{Young2004}
{Young}, P.~R. 2004, \aap, 417, 785

\bibitem[{{Young} {et~al.}(2009){Young}, {Watanabe}, {Hara}, \&
  {Mariska}}]{Young2009}
{Young}, P.~R., {Watanabe}, T., {Hara}, H., \& {Mariska}, J.~T. 2009, \aap,
  495, 587

\end{thebibliography}

\end{document}